# Software-defined subcarrier wave quantum networking operated by OpenFlow protocol

Vladimir V. Chistyakov, Oleg L. Sadov, Artur B. Vasiliev, Vladimir I. Egorov,
Mikhail V. Kompaniets, Pyotr V. Fedchenkov, Oleg I. Lazo, Andrey E. Shevel, Nikita V. Buldakov,
Artur V. Gleim, Sergei E. Khoruzhnikov

*Abstract*— Future practical implementation of secure quantum communication technology in a multiuser network environment would require automatic monitoring of the optical link condition and quantum system parameters, along with adjusting them in real time according to protocol restrictions. We demonstrate how software-defined networking (SDN) paradigm can be used for addressing this problem on the example of a subcarrier wave quantum communication system, which is promising for network applications. We propose a novel approach to dynamic quantum network routing and communication security based on SDN OpenFlow protocol. SDN-operated dynamic switching between different data encryption methods, quantum or classic, will enable to organize virtual encoding channels with an ability to set the required Quality of Service level for security/bandwidth. These developments will further increase the feasibility of quantum technologies from the network perspective and contribute to bringing them to industrial scale.

## I. Introduction

Quantum communication (QC) technologies exploit unique properties of quantum objects to distribute symmetric keys [1], digital signatures [2] or quantum computation data [3] through optical network links with physically ensured security. During the last decade, heroic effort has been put into shifting from experimental point-to-point QC links to multiuser quantum networks functioning in real life environment [4,5]. As a result, quantum networks were successfully launched over the world, and new types of QC devices, promising from networking perspective, were developed. Among them stand subcarrier wave (SCW) QC systems [6-9], most valuable feature of which is exceptionally efficient use of the quantum channel bandwidth and capability of signal multiplexing by adding independent sets of quantum subcarriers to the same carrier wave [8,9]. As shown in [9], in principle this approach allows to increase spectral efficiency of quantum channels by an order of magnitude. It makes SCW QC systems perfect candidates as backbone of multiuser quantum networks.

Rapid development of quantum networking has raised a question of efficient network and key management [4]. It was recently proposed [10-12] that adopting newly developed methods of software defined networking (SDN) in QC networks can increase their flexibility and practicality in several ways. The SDN paradigm separates data transfer from management, leaving the latter to centralized software control plane. It is therefore beneficial in terms of connecting different QC architectures and other network devices in a unified infrastructure. Moreover, as we discuss in this paper, since SDN network controllers can collect operation data from all connected devices, the controller is able to construct a full real-time network map and automatically react to any change of status using a predefined set of instructions. This applies to both network level (e.g. re-routing) and device level (e.g. adjust QC source mean photon number). SDN controller may also be used for automatized switching between quantum and classical encryption methods based on key availability and required Quality of Service (QoS).

On the other hand, QC is promising for protecting SDN control plane data. Since the SDN controller manages the entire network, securing its commands from eavesdropping in some cases can be even more important than protecting the user data. There has been some recent progress in this area: for instance, quantum key distribution was demonstrated for securing network function virtualization in a model SDN network [12].

In this work, we propose how the SDN paradigm can be applied to dynamically operating individual QC systems and networks on a practical example of SCW QC and OpenFlow protocol. This paper is organized as follows. In Section 2, we propose how SDN can be applied to automatically adjusting QC system parameters on a practical example of a SCW QC system. We show which QC components on different levels can be automatically operated based on network data and how it can increase the practicality of QC systems. In Section 3 we describe SDN QC node structure and propose how SDN may

This work was financially supported by Government of Russian Federation, Grant 074-U01 and by the Ministry of Education and Science of Russian Federation: agreement № 14.578.21.0112 (RFMEFI57815X0112), and agreement № 03.G25.31.0229.

V. V. Chistyakov, A. B. Vasiliev, V. I. Egorov, N. V. Buldakov, A. V. Gleim are with the Quantum Information and Communication Laboratory, Department of Photonics and Optical Information Technology, ITMO University, 199034 Kadestkaya line 3b, Saint Petersburg, Russia (e-mail: pspuser@mail.ru, fotonartur@gmail.com, viegorov@corp.ifmo.ru, k101@outlook.c aglejm@yandex.ru).

O. L. Sadov, M. V. Kompaniets, P. V. Fedchenkov, O I. Lazo, A. E. Shevel, S. E. Khoruzhnikov are with the Department of Network and Cloud Technologies, ITMO University, 199034 Birzhevaya line 14, Saint Petersburg, Russia be used for setting data encryption policies based on QC link status. In Section 4 we apply the SDN paradigm to a quantum network as a whole, using OpenFlow protocol for orchestrating quantum routing based on the optical links parameters. Finally, in Section 5 we present the experimental results obtained on an SDN QC testbed. Section 5 summarizes the results obtained in this work.

## II. SOFTWARE-DEFINED OPERATION OF QC DEVICES

In this section, we show how QC system components can be operated based on network data, increasing the practicality of QC setups. Presently the parameters of QC systems are often set during device installation and remain constant in course of operation, and any change in protocol or link status requires manual operations from the administrator. While this is not an obstacle for experimental work or in small-scale networks, future metropolitan and global quantum networks would benefit from automatized control.

This task can be solved by delegating QC network management functions to SDN controllers. In this approach, the controller gathers information from network devices in real time and issues instructions based on a pre-programmed set of rules, which therefore become common to entire network. While SDN controllers communicate with compatible switches using specialized protocols, such as OpenFlow, they may interact with other types of devices through standard interfaces, for example using REST API (see Section 5).

Figure 1 shows a principal scheme of SCW QC system which contains a module (in our example, FPGA) capable of interacting with SDN controller and change parameters of tunable optical components. The transmitter (Alice) module contains a laser which acts as optical carrier source, an optical phase modulator that generates one or several quantum channels on pairs of spectral sidebands [9] and encodes information in them, and a variable attenuator which controls mean photon number in the quantum channel. Here we omit the details of receiver (Bob) module structure; full description of experimental SCW QC setup is given in [8].

Let us consider several examples of QC control based on SDN. It is known that QC protocol parameters, e.g. mean photon number, need to be optimized depending on external factors such as channel loss to achieve a compromise between unconditional security and higher key rate. Normally these hardware parameters are set during system installation. However, in realistic setups optical channel loss can fluctuate due to heat, bending, vibrations and other external influence, as well as gradually increase over time as the cable oldens. Therefore, link losses should be monitored and mean photon number adjusted accordingly during QC system operation. It can be implemented using reflectometers which regularly measure attenuation in the link and upload data to FPGA, which compares it with threshold values of losses set by administrator. If losses exceed critical value, FPGA send a signal to SDN controller which sends status updates to transmitter modules of all QC systems that use current fiber. After receiving the signal, the respective FPGAs issues commands to programmable lasers and/or attenuators which perform fine optical power tuning. This example demonstrates how a single SDN controller may receive status signals from different points and issue commands to any number of devices, thus greatly simplifying network scaling.

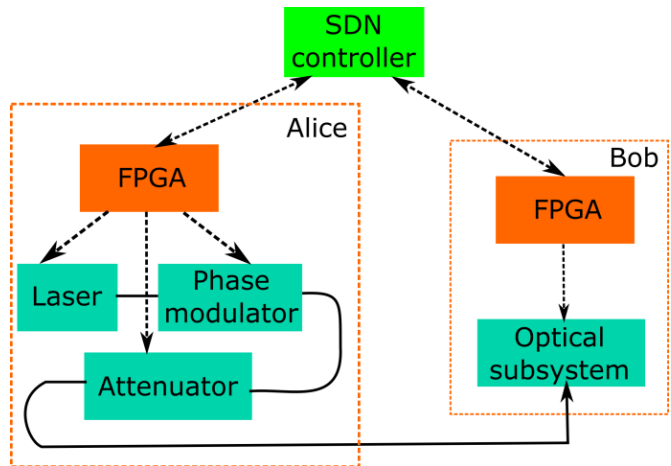

Fig. 1. Principal scheme of SDN-controlled SCW QC system.

Another valuable feature specific for SCW QC method is possibility of adding or dropping independent quantum channels on optical signal sidebands [8] in accordance with current network policies without stopping the system. This can be implemented by external control of phase modulators by FPGA and SDN controller. Let Alice (Fig. 1) generate optical spectrum with a single pair of sidebands (forming one quantum channel) and send it to the Bob module. When an additional Bob module is attached at the end of the same optical link (e.g. by introducing a beam splitter and an optical filter), it sends a request to the controller to establish an additional connection. The controller redirects the request to Alice FPGA which changes modulation parameters in order to generate the second quantum channel on another pair of sidebands around the same carrier wave. This approach will allow increasing spectral efficiency of quantum channels by an order of magnitude using SCW method [9].

Finally, SDN can be used for automatic reconfiguration of quantum network topology. In most QC setups transmitter and receiver modules have different structure and cannot be used bidirectionally. However, a more flexible QC network architecture based on quantum transceivers has been recently proposed in [13]. In this case SDN controller can be used to set status of QC devices (transmitter, receiver or transceiver), based on quantum signal paths through a chain of trusted nodes defined by administrator.

## III. SDN NODE WITH CLASSICAL AND QUANTUM ENCRYPTION

In this section, we describe the architecture of network node connected with a QC system, which uses OpenFlow controller and a compatible SDN switch for data encryption management.

A principal scheme of the suggested OpenFlow-operated QC node structure is shown in Fig. 2. The functions of different network levels are the following. Quantum level is composed of QC systems and their operating software. It is responsible only for quantum information exchange and optical synchronization of Alice and Bob modules. The public channel used for QC protocol data exchange may be implemented by optical devices embedded into the QC modules (as shown in Fig. 2) or realized independently in the transport level. Here we consider that any quantum key processing (sifting, error correction, etc.) is implemented inside the QC modules. Network management level, represented by an SDN controller, operates the routing devices and QC components, monitors links condition, defines optimal signal paths, controls optical switching and data flows. Transport level, realized with an OpenFlow compatible SDN switch, performs physical network routing and data exchange. The switch also serves as an interface to codecs used for data encryption using quantum or classical keys based on controller commands. Finally, the application level represents end-user interfaces and software for network communication.

Let us now describe the data paths in this setup. The SDN controller communicates with switches through OpenFlow protocol and interacts with other connected devices (in our case, QC systems) through specialized software in order to get updates on their status, issue routing commands and set encryption method. Quantum keys are supplied by QC to SDN controller, which sends them to a quantum codec. User data generated on application level is send to a switch, where it is routed to a respective codec according to the controller instructions. Finally, encrypted data is sent to a target node through an optical channel. The quantum keys can also be used by SDN controller for encrypting its commands, thus increasing the whole SDN network robustness against external interference.

Here we consider three codecs (which may be, of course, arbitrary extended): quantum, classical (based on SSL) and transparent (not encrypted). From commutation point of view, they are represented by TCP-services identified by IP-addresses and TCP port numbers for coder at the transmitter and decoder at the receiver. These entrance points are defined in a configuration file of a respective SDN controller application. The controller monitors operational status signals from connected subsystems responsible for certain secure channel functionality (e.g. QC systems). When the subsystem reports malfunction, or its channel is compromised, the controller commands the switch to re-route the data to the next secure channel using a different codec. Since the subsystem reports appears at nearly the same time on transmitter and receiver, data routing is performed simultaneously at both sides.

Another SDN controller application to a QC node is key management. Nowadays even the cutting-edge QC systems demonstrate secure key rates of only 1-2 Mbit/s [8,14], which may be insufficient for encrypting all traffic in a realistic SDN QC network node. A generally accepted solution to this is to use quantum keys for encrypting and distributing traditional symmetric keys, which are in turn frequently updated to maintain the necessary security level. Depending on availability of quantum keys on the server, current amount of traffic and pre-set channel policies, the SDN controller can define the type of codec for each protected channel. For example, if the traffic is relatively low, quantum keys can be used directly for data encoding, thus increasing security to the highest level. When the amount of user data increases, the SDN controller commands the nodes to use quantum keys only for symmetric key updates, and encode information with the latter. Finally, when there are no quantum keys available (e.g. when the QC links are damaged), the controller switches the links to fully classic encryption. Some channels (e.g. used for SDN controller commands transfer) may be marked as "critical", and only quantum encryption is used for them while the keys are available, even if some other, less prioritized links, receive insufficient amounts for maintaining their current security policy.

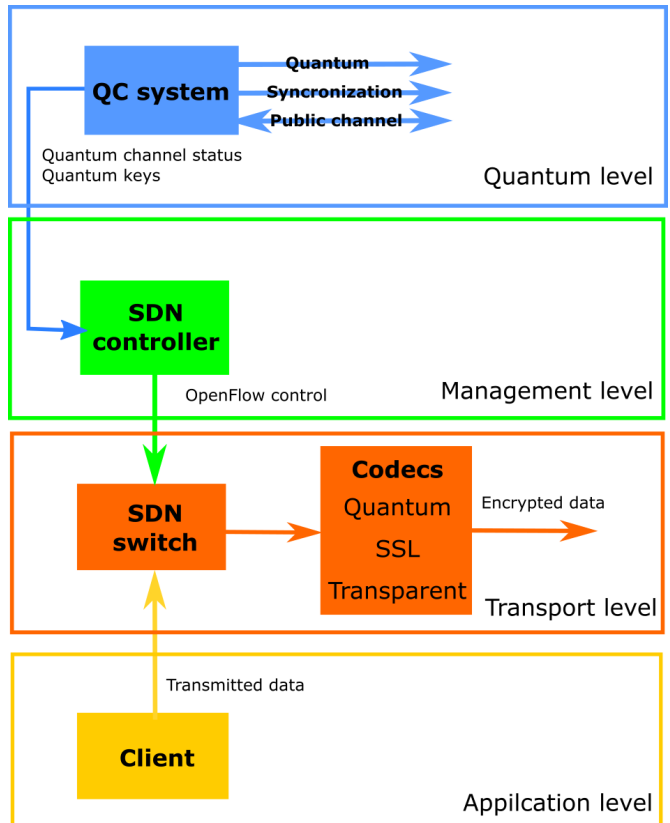

Fig. 2. Principal scheme of OpenFlow-managed QC network node.

Another option is allowing the client to choose between high security/low speed (quantum), high speed/lower security (SSL) and other codecs. Thus, the proposed SDN-operated dynamic switching between different data encryption methods, quantum or classic, will enable to organize virtual encoding channels with an ability to set the desired Quality of Service (QoS) level for security/bandwidth. Our testbed implementation of this approach is discussed in Section 5.

IV. SOFTWARE-DEFINED QC NETWORK WITH DYNAMIC ROUTING

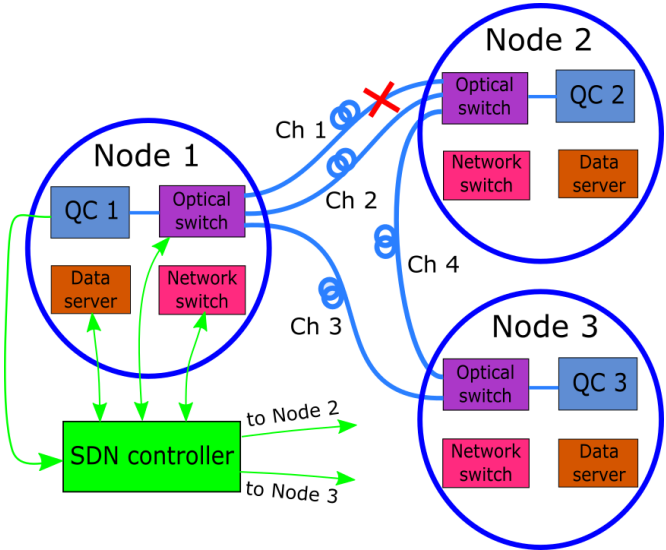

Fig. 3. Software defined quantum communication network structure.

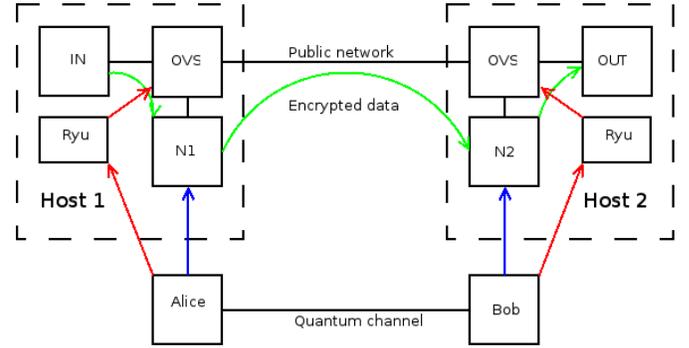

Fig. 4. Principal scheme of SDN-controlled SCW QC system.

Let us discuss how one may apply the SDN paradigm to a quantum network as a whole, using OpenFlow protocol at first for switching between different encryption methods and then for orchestrating classic and quantum routing based on optical links parameters. In a traditional QKD network, routing is static, i.e. all the paths are manually set on switches by a network administrator. In this setup, any change in link status requires constant monitoring, and a problematic QKD session has to be stopped for maintenance. We will show how SDN paradigm overcomes these obstacles, increasing the flexibility and practicality of QC networks.

One should note that in the setup described above (Fig. 2) the transport level was represented by a regular Ethernet switch, which does not support optical signals used in QC technology. In the near future, when the QC networks begin to spread, the problem of quantum signal routing may be solved with other types of devices: SDN operated optical switches, for example, those based on microelectromechanical systems (MEMS) technology. We therefore consider them in the proposed network setup (Fig. 3). In order to allow hop-by-hop quantum key transfer in accordance with "trusted node" QC network paradigm [4], in each node either a pair of QC modules (both Alice and Bob), or a quantum transceiver (e.g. proposed in [13]) is installed. The data servers are used for storing quantum keys and/or received data inside the nodes. Otherwise node structure and data paths are similar to Fig. 2 (some connections are omitted to simplify the image).

In the proposed network architecture, the SDN controller possesses all the information on current status of all supported devices and links between them. This data can be used for automatized path selection in a distributed network based on a number of preset conditions. These conditions may include signs of eavesdropping (e.g. quantum bit error rate (QBER) reaches a critical value), fiber channel damage (no connection), high value of losses, a warning signal from a compromised of damaged node, and others. In all these cases, the SDN controller informs the affected network users and automatically performs switching using backup optical channels. The controller may also be programmed to perform automatic calculation of optimal paths based on real-time monitoring of link parameters. In a large-scale QC network, the difference in total losses for various routes between two nodes can be very substantial, leading to otherwise non-optimal use of bandwidth and lower key rates. Here we consider that a single SDN controller administers the whole network, but in order to make the network it to controller damage several independent controllers with similar functions may be installed at different geographic locations.

Let us see how the network in Fig. 3 reacts to a change in link status. For instance, let users at Nodes 1 and 2 exchange information using the quantum codec. By default, the QC devices in these nodes send quantum signals through channel 1 (Ch 1). If the internal QC diagnostic software reports quantum channel damage (e.g. sifted key rate measured in real time is nearly zero), it signals to the SDN controller. The controller marks Ch 1 as problematic (represented by a red cross in Fig. 2) and sends a command to the optical switch to forward the traffic to another fiber (Ch 2). Now let us consider that each optical link is equipped with an automatized reflectometer which reports values of measured losses to a server located at the node. In this case the SDN controller possesses a regularly updated map of channel loss and can define optimal signal paths. In our model situation, quality of the reserve link (Ch 2) may be poor, so that total loss in links Ch 3 and Ch 4 is much lower, and it is beneficial to re-route the quantum signal through the trusted Node 3. However, from security perspective a direct connection is normally considered safer. This is an example of a network policy that should be defined as a set of rules for the SDN controller by a network administrator. At the same time, if protected channel links (that go through SDN network switches) between Nodes 1 and 2 are not damaged, then encoded information exchange may remain direct, while the keys go through an intermediary Node 3. If Ch1, Ch2 and Ch3 fail at the same time, the controller commands the switch to use SSL codec (see Section 3) for data protection.

V. EXPERIMENTAL SDN QC TEST-BED USING OPENFLOW

In order to demonstrate new network management functions enabled by the proposed approach, we created an experimental SDN operated testbed which uses quantum keys supplied by SCW QC. It demonstrates how a software-defined network with quantum and classical encoding dynamically responds to a change in link status. By default, the SDN controller receives new keys from QC, directs them to quantum codecs and commands the switch to send client data through them. For simplicity, in this demonstration the quantum codec perform exclusive OR (XOR) operation on the data and the keys, thus implementing one-time pad encoding method. When the QC device informs the controller that its quantum channel is compromised (based on QBER level independently measured for each generated key), it switches the transmitter to classic encoding algorithm.

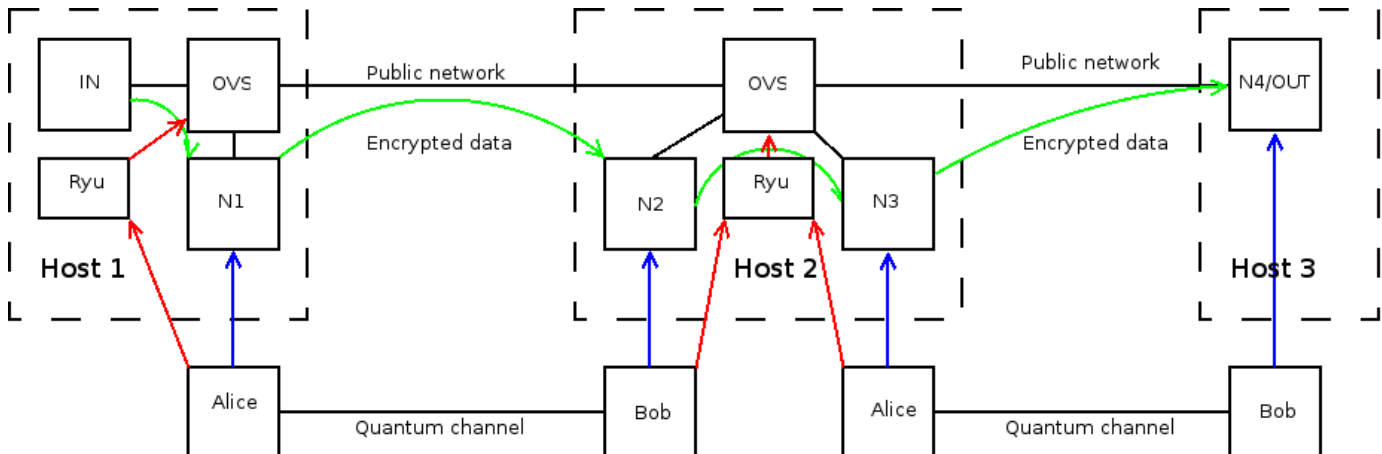

Fig. 5. Example of changing QC channel status and supplying a key in the testbed.

Fig. 6 Principal scheme of SDN-controlled network testbed composed of three nodes and two quantum channels. Quantum key supply is shown in blue lines, user data flow in green lines, control commands in red lines.

The experiments were performed in a three-node network connected by two links in a step-by-step manner, forming two segments. Nodes (or Hosts) 1 and 2 included transmitter and receiver modules of experimental SCW QC setup described in [8] functioning in metropolitan telecommunication infrastructure between two ITMO University buildings in Saint Petersburg, Russia. Quantum key distribution is performed through an underground optical cable consisting of standard SMF-28 fibers. Channel length was 1.63 km, sifted key rate reached 1.06 Mbit/s with mean QBER level 1%. For testing purposes, Node 3 is software emulated using keys generated by the same SCW QC device.

Each segment contains an Alice-Bob pair which supplies quantum keys for encryption/decryption of the data channel. The nodes include a programmable switch, an SDN controller and codecs (marked N1, N2), as shown in Figure 4. Data flow from IN addressed to OUT point (client and server) goes to a programmable switch and is rerouted to active channel on codec N1. Through this channel it arrives to codec N2 where data flow is decrypted and routed to its destination (OUT in Node 2).

In the testbed, we use SDN controllers Ryu and software SDN switches OVS, which communicate over OpenFlow protocol v. 1.3. Mininet was chosen for setting up network infrastructure, because it provides very simple and flexible way for setting up virtual networks with arbitrary topologies.

During normal course of operation, N1 and N2 maintained quantum-encrypted connection. Quantum channel compromising was registered by SCW QC internal software when QBER in sifted key exceeded 11%. In this case the key is discarded and QC malfunction is reported to the controller by Rest API using cURL application. As seen from the first line in Fig.5, the QC system sends a request to SDN controller using GET method, addressing channel 1 of OVS switch on host 192.168.0.76 with parameter «qchannel» and status 0. In the testbed, the SDN controller is configured to respond to zero-status requests by initiating traffic rerouting in addressed channel to SSL codec implemented in N1 by stunnel application. As an alternative, unencrypted connection (transparent codec) could also be established for testing purposes.

When QC operation is restored, the quantum keys generated by SCW QC are pushed to codecs. This is illustrated in Fig. 5 (lines 2 and 3), where a request is first send to the controller indicating status change from 0 to 1 in channel 1. Then a new quantum key in hexadecimal format is uploaded through POST method using parameter «qkey».

Our three-node testbed (Figure 6) consisted of two segments, the structure of which was similar to described above with a few exceptions: Node 3 did not have any switch, as we were not going to re-route any traffic there; codec N4 was merged with OUT client and connected directly to OVS switch on Node 2 in order to simplify configuration. Communication between Nodes 1 and 3 using quantum keys was organized according to trusted repeater paradigm: data was decoded and encoded again in Node 2 using two sets of keys.

## VI. CONCLUSION

We proposed novel approaches to managing quantum communication devices and networks on different levels by using SDN controllers. We considered how SDN may be utilized for automatically adjusting quantum communication device parameters based on changes in link condition. We have also discussed how SDN paradigm may be used for implementing dynamic routing in a multiuser quantum network. We have also shown how OpenFlow SDN switches can be used to organize virtual encoding channels, which support quantum and classical encoding, with an ability to set the required level for security. This function was demonstrated experimentally on a testbed.

The obtained results are important for effective management of future distributed multiuser quantum networks which should be flexible, dynamically adjustable and require no user interference during their normal course of operation.